\documentclass[11pt,reqno,a4paper]{article}
\usepackage{amssymb,bbold,dsfont}
\usepackage{amstext,amsmath,amsfonts,amsthm}
\theoremstyle{amsart}
\newtheorem{theorem}{Theorem}
\newtheorem{prop}{Proposition}
\newtheorem*{propA}{Proposition A}

\date{February 23, 2012}
\title{\protect\vspace{-15mm}\bf On nonlocal symmetries for the Krichever--Novikov equation}
\author{{\Large Petr Voj\v{c}\'{a}k}\\
Mathematical Institute, Silesian University in Opava,\\ Na Rybn\'{i}\v{c}ku 1, 746 01 Opava, Czech Republic\\
E-mail: {\tt Petr.Vojcak@math.slu.cz}}

\begin{document}
\large

\maketitle
%\keywords{Krichever--Novikov equation, nonlocal symmetries, recursion operators}
\begin{abstract}\protect\vspace{-12mm}
\noindent
We construct new infinite hierarchies of nonlocal symmetries and cosymmetries for the Krichever--Novikov equation using the inverse of the fourth-order recursion operator of the latter.
%Note that no nonlocal symmetries and cosymmetries were known for this equation so far. Moreover, there %exist two known recursion operators for the equation in question and we find the inverse of one of them.  \\

\noindent {\bf PACS} 02.30.Ik, 02.30.Jr

\noindent {\bf Keywords:} Krichever--Novikov equation, nonlocal symmetries, recursion operators
\end{abstract}

The Krichever--Novikov (KN) equation
\begin{equation}
\label{kneq}
u_t=u_{xxx} - \frac{3}{2} \frac{u_{xx}^2}{u_x} + \frac{P(u)}{u_x},
\end{equation}
where $P(u)=u^3+c_1u+c_0$ is a third-order polynomial in the reduced form (i.e., without quadratic term and with the leading coefficient equal to $1$), $c_0,c_1 \in {\Bbb R}$,
has first appeared in \cite{krichnov} in connection with the study of finite-gap solutions of the Kadomtsev--Petviashvili equation which has plenty of physical applications from plasma physics to fluid dynamics, see e.g.\  \cite{novikov} and references therein.
Eq.(\ref{kneq}) is integrable in virtually any reasonable sense: it has infinite hierarchies of local higher symmetries, conservation laws, Hamiltonian and symplectic structures, see e.g.\ \cite{demsok}; it also possesses a zero-curvature representation (see e.g.\
\cite{krichnov, im} and references therein) and a plethora of multisoliton and finite-gap solutions  \cite{krichnov, novikov}. The KN equation is the simplest known integrable one-field elliptic model, see \cite{demsok}.\looseness=-1

It is readily seen that using suitable fractional linear
changes of the dependent variable $u$ we can easily turn (\ref{kneq}) into the other known forms of the KN equation with $P$ being a general third- or fourth-degree polynomial in $u$, cf.\ \cite{demsok}.\looseness=-1

%The KN equation is one of the master equations of soliton theory.
%Indeed,
It is well known, see e.g.\ \cite{heredero, mikhailov}, that the family of partial differential equations of the form
\begin{equation}
\label{fam}
u_t=F(x,u,u_x,u_{xx},u_{xxx})
\end{equation}
includes {\em inter alia} three fundamental integrable equations, namely, the linear equation
\begin{equation*}
%\label{lineq}
u_t=u_{xxx}+\alpha(x)u_x+\beta(x)u,
\end{equation*}
the Korteweg--de Vries (KdV) equation
\begin{equation*}
%\label{KdV}
u_t=u_{xxx}+uu_x,
\end{equation*}
and the KN equation (\ref{kneq}).

In fact, it is conjectured, see e.g.\  \cite{heredero, mikhailov} and references therein, that all integrable equations in the family (\ref{fam}) are related via either a classical (point or contact) transformation or a differential substitution, to the above three fundamental integrable equations. Furthermore, this conjecture is proved, see \cite{mikhailov} and references therein, for a subfamily of (\ref{fam}) of the form
\begin{equation*}
%\label{subfam}
u_t=u_{xxx}+\tilde F(x,u,u_x,u_{xx}).
\end{equation*}

The symmetry $G$ of (\ref{fam}) by definition (see e.g.\ \cite{blaszak, v, serg} and references therein) satisfies the equation
\begin{equation}
\label{symeq}
D_t(G)-F_{\star}(G)=0.
\end{equation}
Here and below we identify the symmetry with its characteristic (cf.\ e.g.\ \cite{blaszak, serg}) and use the so-called Fr\'{e}chet (or directional) derivative of $F$,
\begin{equation}
\label{frech}
F_{\star}=\sum\limits_{k=0}^{n}\frac{\partial F}{\partial u_{kx}}D_x^k,
\end{equation}
where $u_{kx}=\partial^k u/\partial x^k$ and $u_{0x}=u$;
\begin{equation}\label{td}
%\begin{aligned}
D_x=\frac{\partial}{\partial x}+\sum\limits_{k=0}^{\infty}u_{(k+1)x}\frac{\partial}{\partial u_{kx}}\quad\mbox{and}\quad
D_t=\frac{\partial}{\partial t}+\sum\limits_{k=0}^{\infty}D_x^k(F)\frac{\partial}{\partial u_{kx}}
%\end{aligned}
%\end{align*}
\end{equation}
respectively denote total derivatives with respect to $x$ and $t$ by virtue of (\ref{fam}), cf. e.g.\ \cite{serg}.

In addition to local variables $x,t,u, u_{kx}$ the symmetry in (\ref{symeq}) may depend on nonlocal variables in the sense of \cite{v} (geometrically, the latter are local coordinates in the fibre of a covering over (\ref{fam})); roughly speaking, the nonlocal variables are understood as pseudopotentials for (\ref{fam})). If this  is the case, it is tacitly assumed that the total derivatives (e.g.\ in (\ref{symeq})) are properly extended to the nonlocal variables. Note that the nonlocal symmetries in this sense, i.e., solutions of (\ref{symeq}) depending on nonlocal variables, are referred to as shadows of nonlocal symmetries in the terminology of \cite{v}.\looseness=-1

The usual way to construct new symmetries from a known symmetry $G$ is to act on $G$ by a recursion operator $\mathcal{R}$. Nearly all known recursion operators for the equations of type (\ref{fam}) have the form
\begin{equation}
\label{rop}
\mathcal{R}=\sum\limits_{i=0}^ka_iD_x^i+\sum\limits_{j=1}^l G_jD_x^{-1} \circ \gamma_j,
\end{equation}
where $G_j$ and $\gamma_j$ are some fixed symmetries and cosymmetries, see e.g.\  \cite{serg} and references therein. Recall (see e.g.\  \cite{blaszak}) that cosymmetries are solutions of the determining equation which is formally adjoint to the one~for symmetries. If $G_j$ and $\gamma_j$, $j\!=\!1, \ldots, l$, are local functions (i.e.,\! they are smooth functions of $x,t, u$~and~of finitely many derivatives of $u$ with respect to $x$) then we call \cite{mn} recursion operators of the form (\ref{rop}) weakly nonlocal.\looseness=-1

There exist \cite{demsok, sokolov} two weakly nonlocal recursion operators of orders 4 and 6 for (\ref{kneq}),
\begin{equation}\label{r1}\displaystyle
\mathcal{R}_1=D_x^4+a_1D_x^3+a_2D_x^2+a_3D_x+a_4+G_1D_x^{-1}\frac{\delta \rho_1}{\delta u}+u_xD_x^{-1}\frac{\delta \rho_2}{\delta u},
\end{equation}
\begin{equation}\label{r2}
\begin{array}{l}
\displaystyle
{\mathcal R}_2=D_x^6+b_1D_x^5+b_2 D_x^4+b_3 D_x^3+b_4 D_x^2+b_5 D_x+b_6-\frac{1}{2} u_x D_x^{-1} \displaystyle\frac{\delta \rho_3}{\delta u}+G_1 D_x^{-1} \displaystyle\frac{\delta \rho_2}{\delta u}+G_2 D_x^{-1}
\frac{\delta \rho_1}{\delta u}.
\end{array}
%\label{KNR2}
\end{equation}
Here
%$u_{jx}=\frac{\partial u}{\partial x^j}$, and
\[
\displaystyle\frac{\delta}{\delta u}=\sum\limits_{j=0}^{\infty}(-D_x)^j \displaystyle\frac{\partial}{\partial u_{jx}}
\]
denotes the variational derivative,
\[
G_1=u_{xxx} - \frac{3}{2} \frac{u_{xx}^2}{u_x} + \frac{P}{u_x},\]
\[
G_2=u_{5x}-5\frac{u_{4x} u_{xx}}{u_x}-\frac{5}{2}
\frac{u_{xxx}^2}{u_x}+\frac{25}{2} \frac{u_{xxx}
u_{xx}^2}{u_x^2}-\frac{45}{8} \frac{u_{xx}^4}{u_x^3}-\frac{5}{3} P
\frac{u_{xxx}}{u_x^2}+\frac{25}{6} P \frac{u_{xx}^2}{u_x^3}-\frac{5}{3}
P' \frac{u_{xx}}{u_x}-\frac{5}{18} \frac{P^2}{u_x^3}+\frac{5}{9} u_x
P'',\]
\[
a_1=-4 \frac{u_{xx}}{u_x}, \hspace{7mm} a_2=-2 \frac{u_{xxx}}{u_x} + 6 \frac{u_{xx}^2}{u_x^2} - \frac{4}{3} \frac{P}{u_x^2},\]
\[
a_3=-2 \frac{u_{4x}}{u_x} + 8 \frac{u_{xx} u_{xxx}}{u_x^2} - 6 \frac{u_{xx}^3}{u_x^3} + 4 \frac{u_{xx}}{u_x^3}P - \frac{2}{3} \frac{P^{\prime}}{u_x},
\]
\[
a_4=\frac{u_{5x}}{u_x} - 4 \frac{u_{xx} u_{4x}}{u_x^2} - 2 \frac{u_{xxx}^2}{u_x^2} + 8 \frac{u_{xx}^2 u_{xxx}}{u_x^3} - 3 \frac{u_{xx}^4}{u_x^4} + \frac{4}{9} \frac{P^2}{u_x^4}+ \frac{4}{3} \frac{u_{xx}^2}{u_x^4}P - \frac{8}{3} \frac{u_{xx}}{u_x^2} P^{\prime} + \frac{10}{9}P^{\prime \prime},
\]
\[
\rho_1=-\frac{1}{2} \frac{u_{xx}^2}{u_x^2} - \frac{1}{3} \frac{P}{u_x^2}, \hspace{7mm} \rho_2=\frac{1}{2} \frac{u_{xxx}^2}{u_x^2} - \frac{3}{8} \frac{u_{xx}^4}{u_x^4} + \frac{5}{6} \frac{u_{xx}^2}{u_x^4}P + \frac{1}{18} \frac{P^2}{u_x^4} - \frac{5}{9}P^{\prime \prime},
\]
\begin{eqnarray*}
\rho_3&=&\displaystyle \frac{u_{4x}^2}{u_x^2}+3 \frac{u_{xxx}^3}{u_x^3}-\frac{19}{2} \frac{u_{xxx}^2 u_{xx}^2}{u_x^4}+
\frac{7}{3}P \frac{u_{xxx}^2}{u_x^4}+\frac{35}{9} P' \frac{u_{xx}^3}{u_x^4}+\frac{45}{8} \frac{u_{xx}^6}{u_x^6}-
\frac{259}{36} \frac{u_{xx}^4 P}{u_x^6}+\frac{35}{18} P^2 \frac{u_{xx}^2}{u_x^6}\\
&&\displaystyle-\frac{14}{9} P'' \frac{u_{xx}^2}{u_x^2}+\frac{1}{27} \frac{P^3}{u_x^6}-\frac{14}{27} \frac{P'' P}{u_x^2}-\frac{7}{27} \frac{P'^2}{u_x^2}-\frac{14}{9}P^{(IV)}u_x^2,
\end{eqnarray*}
\[
b_1=-6\frac{u_{xx}}{u_x},\qquad b_2=-9 \frac{u_{xxx}}{u_x}-2 \frac{P}{u_x^2}+21 \frac{u_{xx}^2}{u_x^2},
\]
\[
b_3=-11 \frac{u_{4x}}{u_x}+60 \frac{u_{xxx} u_{xx}}{u_x^2}+14 P \frac{u_{xx}}{u_x^3}-57 \frac{u_{xx}^3}{u_x^3}-3\frac{P'}{u_x},
\]
\begin{eqnarray*}
b_4&=&\displaystyle-4 \frac{u_{5x}}{u_x}+38 \frac{u_{4x} u_{xx}}{u_x^2}+22 \frac{u_{xxx}^2}{u_x^2}+99 \frac{u_{xx}^4}{u_x^4}-155 \frac{u_{xxx} u_{xx}^2}{u_x^3}+\frac{34}{3} P \frac{u_{xxx}}{u_x^3}-44 P \frac{u_{xx}^2}{u_x^4} +\frac{4}{3} \frac{P^2}{u_x^4}+12 P' \frac{u_{xx}}{u_x^2}-P'',
\end{eqnarray*}
\begin{eqnarray*}
b_5&=&\displaystyle-2 \frac{u_{6x}}{u_x}+29\frac{u_{4x} u_{xxx}}{u_x^2}+80 P\frac{u_{xx}^3}{u_x^5} +
\frac{23}{3} P' \frac{u_{xxx}}{u_x^2} -104 \frac{u_{xx} u_{xxx}^2}{u_x^3} -70\frac{u_{4x} u_{xx}^2}{u_x^3} +241 \frac{u_{xx}^3 u_{xxx}}{u_x^4}+14 \frac{u_{5x} u_{xx}}{u_x^2}\\
&&+\frac{20}{3}P \frac{u_{4x}}{u_x^3}-\frac{170}{3} P\frac{u_{xx}u_{xxx}}{u_x^4} +
\frac{4}{3} \frac{P' P}{u_x^3}-22P' \frac{u_{xx}^2}{u_x^3} +2P'' \frac{u_{xx}}{u_x} -
\frac{16}{3}P^2 \frac{u_{xx}}{u_x^5}-108\frac{u_{xx}^5}{u_x^5},
\end{eqnarray*}
\begin{eqnarray*}
b_6&=&\frac{u_{7x}}{u_x}-6\frac{u_{xx} u_{6x}}{u_x^2} +\frac{8}{9} P^2 \frac{u_{xx}^2}{u_x^6}-
195 \frac{u_{xxx}^2 u_{xx}^2}{u_x^4}+6 P\frac{u_{xxx}^2}{u_x^4}+\frac{142}{3}P \frac{u_{xx}^4}{u_x^6}+\frac{28}{9}P' P\frac{u_{xx}}{u_x^4}+101 \frac{u_{4x} u_{xxx} u_{xx}}{u_x^3}\\
&&+\frac{34}{3} P \frac{u_{4x} u_{xx}}{u_x^4}-72 \frac{u_{xx}^6}{u_x^6}-\frac{28}{9} P''' u_{xx}+\frac{38}{3} P'' \frac{u_{xx}^2}{u_x^2}-\frac{19}{3} P'\frac{u_{4x}}{u_x^2}-\frac{122}{3} P' \frac{u_{xx}^3}{u_x^4}-10 \frac{u_{4x}^2}{u_x^2}+22 \frac{u_{xxx}^3}{u_x^3}\\
&&-\frac{178}{3} P \frac{u_{xxx} u_{xx}^2}{u_x^5}+\frac{14}{9}P^{(IV)}u_x^2+\frac{113}{3} P' \frac{u_{xxx} u_{xx}}{u_x^3}-\frac{2}{3} P \frac{u_{5x}}{u_x^3}-\frac{17}{3} P'' \frac{u_{xxx}}{u_x}-\frac{4}{3} P^2\frac{u_{xxx}}{u_x^5} -89 \frac{u_{4x} u_{xx}^3}{u_x^4}\\
&&+236 \frac{u_{xxx} u_{xx}^4}{u_x^5}-13 \frac{u_{5x} u_{xxx}}{u_x^2}+25\frac{u_{5x} u_{xx}^2}{u_x^3}-\frac{7}{9}\frac{P'^2}{u_x^2}-\frac{8}{27} \frac{P^3}{u_x^ 6}-\frac{4}{9}\frac{P'' P}{u_x^2}.
\end{eqnarray*}
The recursion operators $\mathcal{R}_i$ are not entirely independent: they satisfy \cite{demsok}
the relation (elliptic curve) %check!
\begin{equation}
\mathcal{R}_2^2=\mathcal{R}_1^3-\phi \mathcal{R}_1-\theta, \label{rel}
\end{equation}
where
$$
\begin{array}{l}
\displaystyle \phi=\frac{16}{27}\Big((P'')^2-2 P''' P'+2 P^{(IV)} P\Big),\\[4mm]
\displaystyle \theta=\frac{128}{243}\Big(-\frac{1}{3}(P'')^3-\frac{3}{2}(P')^2P^{(IV)}+P'P''P'''
+2P^{(IV)}P''P-P(P''')^2\Big).
\end{array}
$$

The above naturally leads {\em inter alia} to the following two open problems. First, it is obvious
(cf.\  \cite{demsok}) that the ratio $\mathcal{R}_3=\mathcal{R}_2 \circ \mathcal{R}_1^{-1}$ is a recursion operator of order two for (\ref{kneq}). However, this operator is not weakly nonlocal in the sense of \cite{mn} and, as it was pointed out in \cite{demsok}, it was unclear how to apply it even to the simplest symmetry $u_x$. Second, for many equations it is possible to obtain nonlocal symmetries by applying their recursion operators to the scaling symmetries, see e.g.\ \cite{oevel}. However, the KN equation (\ref{kneq}) has no scaling symmetry and
no nonlocal symmetries for (\ref{kneq}) were known so far.\looseness=-1

%\pagebreak[4]

We address below both of these problems. To this end
we first introduce the
nonlocal variables $p_i$, $q_i$, $z_i$, $i=1,2$, defined by the following relations
(see the appendix for the motivation of this definition):
\begin{equation}\label{cov}
%\begin{align*}
\begin{aligned}
(p_1)_x&=k_3 p_1^2 +2 k_1 p_1-k_2,\quad & %\hspace{7mm}
(p_1)_t&=l_3 p_1^2+2 l_1 p_1-l_2,\\
(z_1)_x&=-(k_1+p_1 k_3),\quad & (z_1)_t&=-(l_1+p_1 l_3),\\
(q_1)_x&=-k_3\exp(-2z_1),\quad & (q_1)_t&=-l_3\exp(-2z_1),\\
(p_2)_x&=-k_3 p_2^2-2 k_1 p_2+k_2,\qquad & (p_2)_t&=-\left(l_3-m\right)p_2^2
-\left(\frac{4u^2}{3u_x}+2l_1\right)p_2-\frac{2c_1}{3u_x}+l_2,\\
(z_2)_x&=(k_1+p_2 k_3),\quad & (z_2)_t&=l_1+\frac{2u^2}{3u_x}+p_2 \left(l_3-m\right),\\
(q_2)_x&=k_3\exp(-2z_2),\quad & (q_2)_t&=\left(l_3-m\right)\exp(-2z_2).
\end{aligned}
%\end{align*}
\end{equation}
Here
\begin{equation} \label{klm}
\begin{aligned}
k_1&=-\frac{\sqrt{6}(c_1u+2c_0)}{12\sqrt{c_0}u_x},\quad k_2=\frac{\sqrt{6}c_1u}{12\sqrt{c_0}u_x},
\quad k_3=\frac{\sqrt{6}u(4c_0u-c_1^2)}{12c_1\sqrt{c_0}u_x},\quad
m=\frac{2(c_1^2-8c_0u-2c_1u^2)}{3c_1u_x},\\
%k_4&=-k_1, & k_5&=-k_2, & k_6&=-k_3,
l_1&=-\frac{\sqrt{6}}{72\sqrt{c_0} u_x^3}(-6 c_1 u u_x u_{xxx} - 12 c_0 u_x u_{xxx} + 3 c_1 u u_{xx}^2 + 6 c_0 u_{xx}^2 + 12 c_1 u_x^2 u_{xx} + 4 \sqrt{6} \sqrt{c_0} u^2 u_x^2\\
& - 2 c_1 u^4 - 4 c_0 u^3 - 2 c_1^2 u^2 - 6 c_0 c_1 u - 4 c_0^2),\\
l_2&=\frac{\sqrt{6}c_1}{72\sqrt{c_0} u_x^3}(-6 u u_x u_{xxx} + 3 u u_{xx}^2 + 12 u_x^2 u_{xx} + 4 \sqrt{6} \sqrt{c_0} u_x^2 - 2 u^4 - 2 c_1 u^2 - 2 c_0 u),\\
l_3&=-\frac{\sqrt{6}}{72c_1\sqrt{c_0} u_x^3}(-6 c_1^2 u u_x u_{xxx} + 24 c_0 u^2 u_x u_{xxx} + 3 c_1^2 u u_{xx}^2 - 12 c_0 u^2 u_{xx}^2 + 12 c_1^2 u_x^2 u_{xx} - 96 c_0 u u_x^2 u_{xx}\\
%\indent \hspace{3mm}
& + 96 c_0 u_x^4 - 4 \sqrt{6} \sqrt{c_0} c_1^2 u_x^2 + 32 \sqrt{6} c_0^{\frac{3}{2}} u u_x^2 + 8 \sqrt{6} \sqrt{c_0} c_1 u^2 u_x^2 + 8 c_0 u^5 - 2 c_1^2 u^4 + 8 c_0 c_1 u^3 - 2 c_1^3 u^2\\
%\indent \hspace{3mm}
& + 8 c_0^2 u^2 - 2 c_0 c_1^2 u).
\end{aligned}
\end{equation}
%\[
%l_4=-\frac{2u^2}{3u_x}-l_1,\qquad l_5=\frac{2c_1}{3u_x}-l_2,\qquad l_6=\frac{2(c_1^2-2c_1u^2-8c_0u)}{3c_1u_x}-l_3.
%\]

%Note that for instance $(p_i)_x$ and $(p_i)_t$, $i=1,2$, denote the usual derivatives of $p_i$ with respect to $x$ and $t$, respectively.
One can readily verify that the integrability conditions $(p_i)_{xt}=(p_i)_{tx}$ are satisfied by virtue of (\ref{kneq}). The same holds true for nonlocal variables $q_i, z_i$, $i=1,2$. Geometrically, equations (\ref{cov}) define a six-dimensional covering (see e.g.\ \cite{v} and references therein for the relevant definitions) over (\ref{kneq}).  While this covering is of general type studied in \cite{im}, its explicit form does not appear there. Further details can be found in the appendix.\looseness=-1

Now define the quantities $V_i$ and $\gamma_i$, $i=1,2,\ldots,6$, as follows:
\begin{eqnarray*}
V_1 &=& \frac{c_1u+2c_0}{2c_0}+\frac{1}{4c_1c_0}
\sum\limits_{i=1}^2\left[((c_1^2-4c_0u)up_i^2+2c_1(c_1u+2c_0)p_i+c_1^2u)(q_i-1)
\exp(2z_i)+u(c_1^2-4c_0u)p_i\right],\\
V_2&=&-\frac{1}{2c_1^2}\sum\limits_{i=1}^2((c_1^2-4c_0u)up_i^2+2c_1(c_1u+2c_0)p_i+c_1^2u)
\exp(2z_i),\\
V_3 &=& \frac{\sqrt{6}}{8c_1^2\sqrt{c_0}}\sum\limits_{i=1}^2(-1)^{i-1}
((c_1^2-4c_0u)up_i^2+2c_1(c_1u+2c_0)p_i+c_1^2u)\exp(2z_i),
\end{eqnarray*}
\begin{eqnarray*}
V_4 &=& -\frac{\sqrt{6}}{64\sqrt{c_0^3}}\sum\limits_{i=1}^2(-1)^{i-1}\left[((c_1^2-4c_0u)up_i^2+2c_1(c_1u+2c_0)p_i+c_1^2u)(q_i-1)^2\exp(2z_i)\right.\\
&&\left.+u(c_1^2-4c_0u)\exp(-2z_i)+2u(c_1^2-4c_0u)p_i(q_i-1)+2c_1(c_1u+2c_0)q_i\right],\\
V_5 &=& -\frac{1}{32c_0}\sum\limits_{i=1}^2\left[((c_1^2-4c_0u)up_i^2+2c_1(c_1u+2c_0)p_i+c_1^2u)(q_i^2-1)\exp(2z_i)\right.\\
&&\left.+u(c_1^2-4c_0u)\exp(-2z_i)+2u(c_1^2-4c_0u)p_iq_i+2c_1(c_1u+2c_0)q_i\right],\\
V_6 &=& \frac{\sqrt{6}}{16c_1\sqrt{c_0}}\sum\limits_{i=1}^2(-1)^{i-1}\left[((c_1^2-4c_0u)up_i^2+2c_1(c_1u+2c_0)p_i+c_1^2u)q_i\exp(2z_i)+u(c_1^2-4c_0u)p_i\right],
\end{eqnarray*}
\begin{eqnarray*}
\gamma_1 &=& \frac{\sqrt{6}}{16c_1\sqrt{c_0}u_x^3}\sum\limits_{i=1}^2(-1)^{i-1}\left\{\left[((c_1^2-4c_0u)up_i^2+2c_1(c_1u+2c_0)p_i+c_1^2u)u_{xx}\right.\right.\\
&&\left.\left.+((8c_0u-c_1^2)p_i^2-2c_1^2p_i-c_1^2)u_x^2\right]q_i\exp(2z_i)
+\left[(8c_0u-c_1^2)u_x^2+(c_1^2-4c_0u)uu_{xx}\right]p_i\right\},\\
\gamma_2 &=& \frac{\sqrt{6}}{64\sqrt{c_0^3}u_x^3}\sum\limits_{i=1}^2(-1)^{i-1}\left\{\left[((c_1^2-4c_0u)up_i^2+2c_1(c_1u+2c_0)p_i+c_1^2u)u_{xx}\right.\right.\\
&&\left.+((8c_0u-c_1^2)p_i^2-2c_1^2p_i-c_1^2)u_x^2\right](q_i-1)^2\exp(2z_i)
+\left[(8c_0u-c_1^2)u_x^2+(c_1^2-4c_0u)uu_{xx}\right]\exp(-2z_i)\\
&&\left.+2\left[(8c_0u-c_1^2)u_x^2+(c_1^2-4c_0u)uu_{xx}\right]p_i(q_i-1)+2c_1\left[(2c_0+c_1u)u_{xx}-c_1u_x^2\right]q_i\right\},
\end{eqnarray*}
\begin{eqnarray*}
\gamma_3 &=& -\frac{1}{16c_0u_x^3}\sum\limits_{i=1}^2\left\{\left[((c_1^2-4c_0u)up_i^2+2c_1(c_1u+2c_0)p_i+c_1^2u)u_{xx}\right.\right.\\
&&\left.+((8c_0u-c_1^2)p_i^2-2c_1^2p_i-c_1^2)u_x^2\right](q_i^2-1)\exp(2z_i)
%\\&&
+\left[(8c_0u-c_1^2)u_x^2+(c_1^2-4c_0u)uu_{xx}\right]\exp(-2z_i)\\
&&\left.+2\left[(8c_0u-c_1^2)u_x^2+(c_1^2-4c_0u)uu_{xx}\right]p_iq_i
+2c_1\left[(2c_0+c_1u)u_{xx}-c_1u_x^2\right]q_i\right\},\\
\gamma_4 &=& \frac{1}{2c_1^2u_x^3}\sum\limits_{i=1}^2\left[((c_1^2-4c_0u)up_i^2+2c_1(c_1u+2c_0)p_i+c_1^2u)u_{xx}+((8c_0u-c_1^2)p_i^2-2c_1^2p_i-c_1^2)u_x^2\right]\exp(2z_i),\\
\gamma_5 &=& \frac{\sqrt{6}}{4c_1^2\sqrt{c_0}u_x^3}\sum\limits_{i=1}^2(-1)^{i-1}
\left[((c_1^2-4c_0u)up_i^2+2c_1(c_1u+2c_0)p_i+c_1^2u)u_{xx}\right.\\
&&\left.+((8c_0u-c_1^2)p_i^2-2c_1^2p_i-c_1^2)u_x^2\right]\exp(2z_i),\\
\gamma_6 &=& \frac{(2c_0+c_1u)u_{xx}-c_1u_x^2}{2c_0u_x^3}+\frac{1}{4c_1c_0u_x^3}\sum\limits_{i=1}^2\left\{\left[((c_1^2-4c_0u)up_i^2+2c_1(c_1u+2c_0)p_i+c_1^2u)u_{xx}\right.\right.\\
&&\left.\left.+((8c_0u-c_1^2)p_i^2-2c_1^2p_i-c_1^2)u_x^2\right](q_i-1)\exp(2z_i)+\left[(8c_0u-c_1^2)u_x^2+(c_1^2-4c_0u)uu_{xx}\right]p_i\right\}.
\end{eqnarray*}

Using the above formulas we arrive at the following assertions
which are readily verified by straightforward (albeit rather tedious) computation.
\begin{prop}\label{p1}
The quantities $V_i$ (resp.\ $\gamma_i$),
$i=1, \ldots, 6$, are nonlocal symmetries (resp.\ nonlocal cosymmetries) for the KN equation (\ref{kneq}).
\end{prop}
Note that in the terminology of \cite{v} the quantities $V_i$ are {\it shadows} of nonlocal symmetries.
Unfortunately we were not able to find a covering which is an extension of (\ref{cov}) and in which these shadows can be lifted to full nonlocal symmetries in the sense of  \cite{v}, and for this reason it was not possible to find the commutators (the Jacobi brackets in the language of \cite{v}) of the symmetries $V_i$. We conjecture, however, that all these commutators vanish when defined in some appropriate setting.

\begin{theorem}\label{t1}
The operator
\begin{equation}\label{iro}\widetilde{\mathcal{R}}=\sum_{i=1}^6 V_i \ D_x^{-1} \circ \gamma_i\end{equation}
is a recursion operator for the KN equation (\ref{kneq}) which is the inverse of $\mathcal{R}_1$, that is,
\begin{equation}\label{roiro}\widetilde{\mathcal{R}} \circ \mathcal{R}_1=\mathcal{R}_1 \circ
\widetilde{\mathcal{R}} = \mathds{1}.
\end{equation}
%thus $\widetilde{\mathcal{R}}$ is the inverse of $\mathcal{R}_1$.
\end{theorem}
The explicit form of $\widetilde{\mathcal{R}}$ was found using the technique from \cite{marvan}, see the appendix for details. As we can see, it is possible to write $\widetilde{\mathcal{R}}$ in the standard form, with $D_x^{-1}$ appearing in each term at most once. Nevertheless, the operator $\widetilde{\mathcal{R}}$ is {\em not} weakly nonlocal in the sense of \cite{mn}, as
$V_j$ and $\gamma_j$ are not local themselves, cf.\ e.g.\ \cite{marvan, sd} for other examples of this kind. This very fact ensures that (\ref{iro}) is inverse to $\mathcal{R}_1$ even though one would naively expect that such an inverse is of order minus four rather than minus one, see the discussion in  \cite{serg05}. As for $V_j$ and $\gamma_j$, what we have here is just another example of nonlocal symmetries and cosymmetries arising from the intrinsic structure of the recursion operator rather than from its action on the scaling symmetry, cf.\ e.g.\ \cite{serg05}.\looseness=-1

Notice that for the correct computation of the action of $\widetilde{\mathcal{R}}$ one should use its so-called Guthrie's form, which is given in Proposition~A in the appendix. %However, the resulting expressions are
%extremely cumbersome and for this reason are omitted in the present paper.\looseness=-1

Using Theorem~\ref{t1} we can enhance the result of Proposition~\ref{p1} as follows:
\begin{prop}\label{p2}
The quantities $V_{i}^{(j)}=\widetilde{\mathcal{R}}^j (V_i)$ (resp.\ $(\widetilde{\mathcal{R}}^{\star})^j(\gamma_i)$),
$i=1, \ldots, 6$, are nonlocal symmetries (resp.\ nonlocal cosymmetries) for the KN equation (\ref{kneq})
for all $j=0,1,2,\dots$.
\end{prop}
Here $\widetilde{\mathcal{R}}^{\star}=-\sum_{i=1}^6  \gamma_i \ D_x^{-1} \circ V_i$ is the formal adjoint of $\widetilde{\mathcal{R}}$.

In a similar fashion, it is possible to construct two hierarchies of highly nonlocal Hamiltonian structures for (\ref{kneq}) of the form $\widetilde{\mathcal{R}}^j \circ \mathcal{H}_i$, $j=1,2,\dots$, $i=0,2$, where
$\mathcal{H}_0=u_x D_x^{-1}\circ u_x$ and $\mathcal{H}_2=\mathcal{R}_2\circ \mathcal{H}_0$, cf.\ \cite{demsok, sokolov}.

It is an interesting open problem to find out whether there exist nonlocal conservation laws for (\ref{kneq}) (or, more precisely, for the extended system
(\ref{kneq})+(\ref{cov})) associated with the cosymmetries $(\widetilde{\mathcal{R}}^{\star})^j(\gamma_i)$.

Thus, we have constructed new infinite hierarchies of nonlocal symmetries and cosymmetries for (\ref{kneq}).
It is readily verified that, as $V_j$ do not depend explicitly on $x$ and $t$, they commute with $u_x$ and $G_1$.
In turn, as $\mathcal{R}_1$, and hence $\widetilde{\mathcal{R}}$, are hereditary, this implies that
$V_{i}^{(j)}$ commute with the well-known (see e.g.\ \cite{demsok}) local symmetries $\mathcal{R}_1^k(u_x)$ and $\mathcal{R}_1^l (G_1)$ for all $j,k,l=0,1,2,\dots$ and $i=1,\dots,6$.
%check!

Now consider the action of $\widetilde{\mathcal{R}}$ on the known local symmetries of (\ref{kneq}).
First of all, the action on $\mathcal{R}_1^k(u_x)$ and $\mathcal{R}_1^l (G_1)$ for $k,l=1,2,\dots$ is obvious from (\ref{roiro}). Next, $\widetilde{\mathcal{R}}(u_x)$ and
$\widetilde{\mathcal{R}}(G_1)$ turn out to lie in the span of $V_j$, so no new symmetries arise here.
As the action of $D_x^{-1}$ is, roughly speaking, defined only up to the
addition of an arbitrary constant, see e.g.\ \cite{serg} and references therein for details,
it is natural to consider the symmetries resulting from the action of $\widetilde{\mathcal{R}}$ modulo the linear subspace $\mathcal{V}$ spanned by $V_i$, $i=1,\dots,6$. Then $\widetilde{\mathcal{R}}(u_x)$ and $\widetilde{\mathcal{R}}(G_1)$ fall into the equivalence class of zero symmetry, and subsequent action of $\mathcal{R}_2$ on this trivial symmetry produces nothing of interest. Conversely, it is readily checked that the action of $\mathcal{R}_1$ on $V_j$ yields the symmetries that are linear combinations of $u_x$ and $G_1$ and thus again belong to the equivalence class of zero symmetry.\looseness=-1

On the other hand, it is easily verified that the action of $\mathcal{R}_2$ on $V_j$ also gives nothing new, as the resulting symmetries again lie in the span $\mathcal{V}$ of $V_j$:
\begin{equation}\label{r2u}
\begin{aligned}
\mathcal{R}_2(V_1)&=\frac{32}{9}c_1 V_3-\frac{64}{9}V_6,\quad & %\\[1mm]
\mathcal{R}_2(V_2)&=\frac{64}{9}c_0 V_3,\\[1mm]
\mathcal{R}_2(V_3)&=\frac{8}{3}V_2,\quad & %\\[1mm]
\mathcal{R}_2(V_4)&=-\frac{4}{3}c_1 V_1-\frac{16}{3}V_5,\\[1mm]
\mathcal{R}_2(V_5)&=-\frac{8}{9}c_1^2 V_3-\frac{32}{9}c_0 V_4+\frac{16}{9}c_1 V_6,\quad & %\\[1mm]
\mathcal{R}_2(V_6)&=-\frac{8}{3}c_0 V_1+\frac{4}{3}c_1 V_2.
\end{aligned}
\end{equation}

Strictly speaking, the above formulas hold modulo the linear subspace spanned by $u_x$ and $G_i$, $i=1,2$, cf.\ the discussion of the action of $D_x^{-1}$ in the previous paragraph. It is now clear that the repeated application of  $\mathcal{R}_2$ to $V_j$ also produces no new symmetries, as one can readily infer from (\ref{r2u}) or from (\ref{rel}).

Quite obviously, the above discussion entirely settles the issue of the form of symmetry $\mathcal{R}_2 \circ \mathcal{R}_1^{-1}(u_x)$ raised in \cite{demsok}. Moreover, the explicit form of the recursion operator $\mathcal{R}_3=\mathcal{R}_2 \circ \mathcal{R}_1^{-1}$ can now be readily found using Theorem~\ref{t1}
in conjunction with (\ref{r2}); likewise, the Guthrie form (see appendix)
of $\mathcal{R}_3$ can be easily extracted from that of $\mathcal{R}_2$ and Proposition A.

As a final remark, note that
it would be very interesting to figure out how the symmetries $V_{i}^{(j)}$
act on the known solutions (in particular, multisoliton and finite-gap ones)
of the KN equation, cf.\ e.g.\ \cite{gos} and references therein,
where the action of nonlocal symmetries on solutions is studied for the case of KdV and KP equations,
and to find the explicit form of solutions invariant under $V_{i}^{(j)}$, cf.\ e.g.\  \cite{reyes} for the case of Kaup-- Kupershmidt equation and \cite{reyes2} for the Korteweg--de Vries,
Camassa--Holm and Hunter--Saxton equations.

\subsection*{Acknowledgments}
The author is sincerely grateful to Dr. A. Sergyeyev for his highly stimulating attention to this work. It is my pleasure to thank the referees for useful suggestions. This research was supported by Silesian university in Opava
under the student grant SGS/18/2010 and by the Ministry of Education,
Youth and Sports of the Czech Republic under the grant MSM 4781305904.

%%\appendix{Appendix: Proof of Theorem 1}
\section*{Appendix: Proof of Theorem 1}

Here we outline the proof of Theorem \ref{t1}. To this end it is helpful to rewrite the recursion operator $\mathcal{R}_1$ in the so-called Guthrie form, i.e., as a B\"{a}cklund auto-transformation for (\ref{symeq}) with defined by (\ref{kneq}), cf.\ e.g.\ \cite{v, gut, marvan, ms, ms2, serg}). Technically this amounts, roughly speaking, to writing the result of action of the recursion operator (in our case, $\mathcal{R}_1$)
on an arbitrary symmetry $U$ of (\ref{kneq}) and defining the nonlocal quantities arising
in this action as a covering in the sense of \cite{v}. For $\mathcal{R}_1$ this yields
\begin{equation}
\label{rogut}
\widetilde{U}=\mathcal{R}_1(U)=D_x^4(U)+a_1 D_x^3(U)+a_2 D_x^2(U)+a_3 D_x(U)+a_4 U+G_1 W_1+u_x W_2.
\tag{A.1}
\end{equation}
Here the nonlocal variables $W_j$, $j=1,2$ are defined by the relations (see e.g.\ (11) in \cite{serg})):
\begin{equation}\label{nvw}\tag{A.2}
%\begin{align*}
\begin{aligned}
(W_j)_x&=\frac{\delta \rho_j}{\delta u}U,\quad &
(W_j)_t&=\sum\limits_{k=1}^3\sum\limits_{m=0}^{k-1}(-D_x)^m\left(\frac{\partial G_1}{\partial u_{kx}}\frac{\delta \rho_j}{\delta u}\right)D_x^{k-m-1}(U).
\end{aligned}
%\end{align*}
\end{equation}
Recall that $G_1$ denotes the right-hand side of (\ref{kneq}) and $u_{kx}=\partial^{k}u/\partial x^k$. Further define the quantities $U_i$, $i=1,\ldots, 3$, by the formulas
\begin{equation}\label{Ui}\tag{A.3}
%\begin{align*}
\begin{aligned}
D_x(U)&=U_1,\\
D_x(U_1)&=U_2,\\
D_x(U_2)&=U_3,\\
D_x(U_3)&=\widetilde{U}-a_1U_3-a_2U_2-a_3U_1-a_4U-G_1W_1-u_xW_2,\\
D_t(U_i)&=D_x^i((G_1)_{\star}(U)).
\end{aligned}
%\end{align*}
\end{equation}
Let $\vec{\Psi}=(U,U_1,U_2,U_3,W_1,W_2)^T$. Then there exist  matrices $A,B,M_1,M_2$ such that
\begin{equation}\label{ZCR}\tag{A.5}
\begin{aligned}
D_x(\vec{\Psi})&=A\vec{\Psi}+M_1\widetilde{U},\quad &
D_t(\vec{\Psi})&=B\vec{\Psi}+M_2\widetilde{U},
\end{aligned}
%\end{align*}
\end{equation}
where we tacitly assume that the action of total derivatives was extended to the nonlocal variables $W_i$ using (\ref{nvw}), cf.\ the discussion after (\ref{td}). Note that in fact $M_2$ is a matrix-valued differential operator.

It is readily seen that
$$A=\left(\begin{array}{@{}cccccc@{}} 0 & 1 & 0 & 0 & 0 & 0 \\ 0 & 0 &
1 & 0 & 0 & 0 \\ 0 & 0 & 0 & 1 & 0 & 0 \\ -a_4 & -a_3 &
-a_2 & -a_1 & -G_1 & -u_x \\ \vspace{2mm} \displaystyle  \frac{\delta \rho_1}{\delta u} & 0 & 0 &
0 & 0 & 0 \\ \displaystyle  \frac{\delta \rho_2}{\delta u}& 0 & 0 & 0 & 0 & 0 \end{array}\right), \hspace{2mm} M_1=\left(\begin{array}{@{}c@{}} 0 \\ 0 \\ 0 \\ 1 \\ 0 \\ 0
\end{array}\right), \hspace{2mm}
M_2=\left(\begin{array}{@{}c@{}} 0 \\ 1 \\
\vspace{2mm}  \displaystyle D_x+\frac{u_{xx}}{u_x} \\
\displaystyle D_{x}^2+\frac{u_{xx}}{u_x}D_x+ \frac{6 u_x u_{xxx} + 3 u_{xx}^2 + 2 P}{6u_x^2} \\
0 \\ 0 \end{array}\right).$$
The matrix $B$ is rather involved, so its explicit form is omitted. The matrices $A$ and $B$ form a zero-curvature representation (ZCR), see e.g.\ \cite{novikov} for the definition and basic properties of the  latter, for (\ref{kneq}). The entries of $A$ depend on $u,u_x,\ldots,u_{6x}$, but we shall use a gauge-equivalent ZCR, cf.\ e.g. \cite{mar2} and references therein, which is associated with the matrices
\begin{equation}\label{gt}\tag{A.6}
\begin{aligned}
\widetilde{A}=D_x(S^{(0)})(S^{(0)})^{-1}+S^{(0)}A(S^{(0)})^{-1}, \quad & \widetilde{B}=D_t(S^{(0)})(S^{(0)})^{-1}+S^{(0)}B(S^{(0)})^{-1}.
\end{aligned}
%\end{align*}
\end{equation}
Here
$$S^{(0)}=\left(\begin{array}{@{}cccccc@{}} \vspace{1mm}
S^{(0)}_{11} &
\displaystyle -\frac{u}{u_x} &\displaystyle  \frac{1}{4} \frac{u^2}{u_x^2} & 0 &
\displaystyle -\frac{1}{4} \frac{u^2}{u_x} & 0 \\ \vspace{1mm}
S^{(0)}_{21}&
\displaystyle -\frac{1}{u_x} &\displaystyle  \frac{1}{2} \frac{u}{u_x^2} & 0 &
\displaystyle -\frac{1}{2} \frac{u}{u_x} & 0 \\ \vspace{1mm}
S^{(0)}_{31}& S^{(0)}_{32} &
S^{(0)}_{33}&
\displaystyle \frac{1}{2} \frac{u^2 u_{xx}}{u_x^3} - \frac{u}{u_x} &
S^{(0)}_{35}&
\displaystyle -\frac{1}{2} \frac{u^2}{u_x} \\ \vspace{1mm}
S^{(0)}_{41} &
S^{(0)}_{42} &
S^{(0)}_{43} &
\displaystyle \frac{u u_{xx}}{u_x^3} - \frac{1}{u_x} &
S^{(0)}_{45}&
\displaystyle -\frac{u}{u_x} \\ \vspace{2mm}
S^{(0)}_{51} &
0 & \displaystyle \frac{1}{u_x^2} & 0 &\displaystyle  -\frac{1}{u_x} & 0 \\
S^{(0)}_{61} &
S^{(0)}_{62} &
\displaystyle -\frac{u_{xxx}}{u_x^3} - \frac{u_{xx}^2}{u_x^4} &
\displaystyle \frac{u_{xx}}{u_x^3} &
\displaystyle \frac{1}{2} \frac{u_{xx}^2}{u_x^3} - \frac{1}{3u_x^3}P &\displaystyle  -\frac{1}{u_x} \end{array}\right),$$\\
where
\[
S^{(0)}_{11}=-\frac{1}{4} \frac{u^2 u_{xxx}}{u_x^3} + \frac{1}{4} \frac{u^2 u_{xx}^2}{u_x^4} - \frac{1}{6} \frac{u^2}{u_x^4}P + 1,
\hspace{7mm} S^{(0)}_{21}=-\frac{1}{2} \frac{u u_{xxx}}{u_x^3} + \frac{1}{2} \frac{u u_{xx}^2}{u_x^4} - \frac{1}{3} \frac{u}{u_x^4}P,
\]
\begin{eqnarray*}
S^{(0)}_{31}&=&\frac{2u^2 u_x u_{xxx} - 4 u^2 u_{xx}^2 - 2u u_x^2 u_{xx} + 2u_x^4}{3u_x^6}P+\frac{1}{6} \frac{u(5 u u_{xx} - 2 u_x^2)}{u_x^4}P^{\prime}-\frac{5}{18} \frac{u^2}{u_x^2}P^{\prime \prime}-\frac{1}{2} \frac{u^2 u_{5x}}{u_x^3} + 2 \frac{u^2 u_{xx} u_{4x}}{u_x^4}\\
&+& \frac{3}{2} \frac{u^2 u_{xxx}^2}{u_x^4} - \frac{(5 u^2 u_{xx}^2 + u u_x^2 u_{xx} - u_x^4) u_{xxx}}{u_x^5} + \frac{(u u_{xx} + u_x^2)(2 u u_{xx} - u_x^2) u_{xx}^2}{u_x^6},
\end{eqnarray*}
\[
S^{(0)}_{32}=-\frac{2u(u u_{xx} - u_x^2)}{3u_x^5}P+\frac{u^2}{6u_x^3}P^{\prime}+\frac{1}{2} \frac{u^2 u_{xxxx}}{u_x^3} + \frac{1}{2} \frac{(-3 u u_{xx} + 2 u_x^2) u u_{xxx}}{u_x^4} - \frac{(u u_{xx} + u_x^2)(-u u_{xx} + 2 u_x^2) u_{xx}}{u_x^5},
\]
\[
S^{(0)}_{33}=-\frac{1}{2} \frac{u^2 u_{xxx}}{u_x^3} - \frac{1}{2} \frac{u^2 u_{xx}^2}{u_x^4} + 2 \frac{u u_{xx}}{u_x^2} + 1,
\hspace{7mm} S^{(0)}_{35}=\frac{1}{4} \frac{u^2 u_{xx}^2}{u_x^3} - \frac{u u_{xx}}{u_x} + u_x - \frac{1}{6} \frac{P u^2}{u_x^3},
\]
\begin{eqnarray*}
S^{(0)}_{41}&=&\frac{4 u u_x u_{xxx} -8 u u_{xx}^2 -2 u_x^2 u_{xx}}{3u_x^6}P+\frac{5 u u_{xx} - u_x^2}{3u_x^4}P^{\prime}-\frac{5u}{9u_x^2}P^{\prime \prime}-\frac{u u_{5x}}{u_x^3} + 4 \frac{u u_{xx} u_{4x}}{u_x^4} + 3 \frac{u u_{xxx}^2}{u_x^4}\\
&& - \frac{(10 u u_{xx} + u_x^2) u_{xx} u_{xxx}}{u_x^5} + \frac{(4 u u_{xx} + u_x^2) u_{xx}^3}{u_x^6},
\end{eqnarray*}
\[
S^{(0)}_{42}=\frac{1}{3} \frac{-4 u u_{xx} + 2 u_x^2}{u_x^5} P+ \frac{1}{3} \frac{u}{u_x^3}P^{\prime}+\frac{u u_{4x}}{u_x^3} + \frac{(-3 u u_{xx} + u_x^2) u_{xxx}}{u_x^4} - \frac{(-2 u u_{xx} + u_x^2) u_{xx}^2}{u_x^5},
\]
\[
S^{(0)}_{43}=-\frac{u u_{xxx}}{u_x^3} - \frac{u u_{xx}^2}{u_x^4} + 2 \frac{u_{xx}}{u_x^2}, \hspace{7mm} S^{(0)}_{45}=\frac{u u_{xx}^2}{2u_x^3} - \frac{u_{xx}}{u_x} - \frac{u}{3u_x^3}P, \hspace{7mm} S^{(0)}_{51}=-\frac{u_{xxx}}{u_x^3} + \frac{u_{xx}^2}{u_x^4} - \frac{2}{3u_x^4}P,
\]
\[
S^{(0)}_{61}=\frac{4u_x u_{xxx} - 8 u_{xx}^2}{3u_x^6}P+\frac{5}{3} \frac{u_{xx} }{u_x^4}P^{\prime}-\frac{5}{9u_x^2}P^{\prime \prime}-\frac{u_{5x}}{u_x^3} + 4 \frac{u_{xx} u_{4x}}{u_x^4} + 3 \frac{u_{xxx}^2}{u_x^4} - 10 \frac{u_{xx}^2 u_{xxx}}{u_x^5} + 4 \frac{u_{xx}^4}{u_x^6},
\]
\[
S^{(0)}_{62}=-\frac{4}{3} \frac{u_{xx}}{u_x^5}P+\frac{1}{3u_x^3}P^{\prime}+\frac{u_{4x}}{u_x^3} - 3 \frac{u_{xx} u_{xxx}}{u_x^4} + 2 \frac{u_{xx}^3}{u_x^5}.
\]

%\pagebreak[4]

The entries of the matrix $\widetilde{A}$ depend only on $u$ and $u_x$. The ZCR associated with the matrices $\widetilde{A}$ and $\widetilde{B}$ lives in the Lie algebra $\mathcal{A}=\mathfrak{sl}_2\bigoplus \mathfrak{sl}_2$, which is generated by
\begin{equation}\label{rsd}\tag{A.7}
%\begin{align*}
\begin{aligned}
\hspace*{-7mm}
X_1&=-\frac{\sqrt{6}}{2\sqrt{c_0}}Q_1+\frac{\sqrt{6c_0}}{c_1}Q_2
+\frac{c_1\sqrt{6}}{4\sqrt{c_0}}Q_3+\frac{3}{2c_1}Q_4+\frac{3c_1}{16c_0}Q_5,\quad &
Y_1&=\frac{c_1\sqrt{6}}{4\sqrt{c_0}}Q_3-\frac{3c_1}{16c_0}Q_5,\\
H_1&=-\frac{\sqrt{6}}{2\sqrt{c_0}}Q_1+\frac{3c_1}{8c_0}Q_5-\frac{3}{2}Q_6, \\
X_2&=\frac{\sqrt{6}}{2\sqrt{c_0}}Q_1-\frac{\sqrt{6c_0}}{c_1}Q_2
-\frac{c_1\sqrt{6}}{4\sqrt{c_0}}Q_3+\frac{3}{2c_1}Q_4+\frac{3c_1}{16c_0}Q_5,\quad &
Y_2&=-\frac{c_1\sqrt{6}}{4\sqrt{c_0}}Q_3-\frac{3c_1}{16c_0}Q_5,\\
H_2&=\frac{\sqrt{6}}{2\sqrt{c_0}}Q_1+\frac{3c_1}{8c_0}Q_5-\frac{3}{2}Q_6,
\end{aligned}
%\end{align*}
\end{equation}
where
$$Q_1=\left(\begin{array}{@{}cccccc@{}} 0 & 0 & 0 & 0 & 0 & 0 \\ 0 & 0 &
-\frac{1}{2} & 0 & 0 & 0 \\ -\frac{2}{3} c_1 &
-\frac{4}{3} c_0 & 0 & 0 & 0 & 0 \\ 0 & 0 & 0 & 0 &
-\frac{2}{3} c_0 & 0 \\ 0 & 0 & 0 & -1 & 0 & 0 \\ 0 & 0 & 0 &
0 & \frac{1}{3} c_1 & 0 \end{array}\right), \hspace{5mm}
Q_2=\left(\begin{array}{@{}cccccc@{}} 0 & 0 & -\frac{1}{2} & 0 & 0 & 0 \\
0 & 0 & 0 & 0 & 0 & 0 \\ 0 & -\frac{2}{3} c_1 & 0 & 0 & 0 & 0 \\
\frac{4}{3} & 0 & 0 & 0 & -\frac{1}{3} c_1 & 0 \\ 0 & 0 & 0 &
0 & 0 & 1 \\ 0 & \frac{4}{3} & 0 & 0 & 0 & 0 \end{array}\right),$$
$$Q_3=\left(\begin{array}{@{}cccccc@{}} 0 & 0 & 0 & \frac{1}{4} & 0 & 0 \\
0 & 0 & 0 & 0 & 0 & \frac{1}{4} \\ -\frac{2}{3} & 0 & 0 & 0 & 0 &
0 \\ 0 & 0 & 0 & 0 & 0 & 0 \\ 0 & 0 & 0 & 0 & 0 & 0 \\ 0 & 0 & 0 &
0 & \frac{1}{3} & 0 \end{array}\right), \hspace{5mm}
Q_4=\left(\begin{array}{@{}cccccc@{}} \frac{2}{3} c_1 &
\frac{4}{3} c_0 & 0 & 0 & 0 & 0 \\ 0 & -\frac{2}{3} c_1 &
0 & 0 & 0 & 0 \\ 0 & 0 & 0 & 0 & 0 & 0 \\ 0 & 0 & 0 &
\frac{2}{3} c_1 & 0 & \frac{4}{3} c_0 \\ \frac{8}{3} & 0 &
0 & 0 & 0 & 0 \\ 0 & 0 & -\frac{4}{3} & 0 & 0 & -\frac{2}{3} c_1
\end{array}\right),$$
$$Q_5=\left(\begin{array}{@{}cccccc@{}} 0 & 0 & 0 & 0 &
-\frac{2}{3} c_0 & 0 \\ -\frac{4}{3} & 0 & 0 & 0 &
\frac{1}{3} c_1 & 0 \\ 0 & 0 & 0 &\frac{2}{3} c_1 & 0 &
\frac{4}{3} c_0 \\ 0 & 0 & 0 & 0 & 0 & 0 \\ 0 & 0 & 0 & 0 & 0 &
0 \\ 0 & 0 & 0 & -\frac{4}{3} & 0 & 0 \end{array}\right), \hspace{5mm}
Q_6=\left(\begin{array}{@{}cccccc@{}} 0 & 0 & 0 & 0 &
-\frac{1}{6} c_1 & 0 \\ 0 & \frac{2}{3} & 0 & 0 & 0 & 0 \\ 0 &
0 & \frac{2}{3} & 0 & 0 & \frac{1}{3} c_1 \\ 0 & 0 & 0 &
-\frac{2}{3} & 0 & 0 \\ 0 & 0 & 0 & 0 & -\frac{2}{3} & 0 \\ 0 & 0 &
0 & 0 & 0 & 0 \end{array}\right).$$\\
Note that $H_1,H_2$ form a Cartan subalgebra $\mathcal{C}$ of $\mathcal{A}$. Furthermore, we have
\begin{equation*}%\label{gezcr}\tag{A.8}
%\begin{align*}
\begin{aligned}
\widetilde{A}&=k_1H_1+k_2X_1+k_3Y_1-k_1H_2-k_2X_2-k_3Y_2,\\
\widetilde{B}&=l_1H_1+l_2X_1+l_3Y_1-\left(l_1+\frac{2u^2}{3u_x}\right)H_2-\left(l_2-\frac{2c_1}{3u_x}\right)X_2-\left(l_3-m\right)Y_2,
\end{aligned}
%\end{align*}
\end{equation*}
where $m,k_j, l_j$, $j=1,2,3$, are given above in (\ref{klm}). In analogy with Example 28 from \cite{mar2}, we define nonlocal variables $p_i,z_i,q_i$, $i=1,2$, by (\ref{cov}). If we now put $\vec{\Phi}=S\vec{\Psi}$, where
\begin{equation}\label{def-s}\tag{A.8}
S=\exp(q_2Y_2)\cdot \exp(z_2H_2) \cdot \exp(p_2X_2) \cdot \exp(q_1Y_1)\cdot \exp(z_1H_1) \cdot \exp(p_1X_1) \cdot S^{(0)},
\end{equation}
then we readily find from (\ref{ZCR}) that
\begin{equation}\label{phieq}\tag{A.9}
\begin{aligned}
D_x(\vec{\Phi})&=SM_1\widetilde{U},\quad &
D_t(\vec{\Phi})&=SM_2\widetilde{U}.
\end{aligned}
%\end{align*}
\end{equation}

Recall that by definition the first component of $\vec{\Psi}$ is $U$. Therefore, the action of recursion operator $\widetilde{\mathcal{R}}=\mathcal{R}_1^{-1}$ in the Guthrie form on a symmetry $\widetilde{U}$ is given by the formula
\begin{equation} \label{inverseR}\tag{A.10}
\widetilde{\mathcal{R}}(\widetilde{U})=(S^{-1}\vec{\Phi})_1,
\end{equation}
where the subscript $1$ indicates the first component and $\vec{\Phi}$ is now assumed to be {\em defined} via (\ref{phieq}) with $S$ given by (\ref{def-s}). Thus, in order to apply $\widetilde{\mathcal{R}}$ to any (possibly nonlocal) symmetry $\widetilde U$ of (\ref{kneq}) we should first compute the vector nonlocal quantity $\vec{\Phi}$ defined by (\ref{phieq}), and then the right-hand side of (\ref{inverseR}) is the sought-for new symmetry $\widetilde{\mathcal{R}}(\widetilde{U})$. Note that the quantities $V_i$, $i=1,\ldots,6$, from Proposition \ref{p1}, whose generic linear combination is the image of zero symmetry under the action of $\widetilde{\mathcal{R}}$, form the first row of the matrix $S^{-1}$.\looseness=-1

Thus, we have the following
\begin{propA}\label{iroprop}
The KN equation (\ref{kneq}) possesses a recursion operator $\widetilde{\mathcal{R}}$ whose action on a
(possibly nonlocal) symmetry $U$ of (\ref{kneq}) has the form
\begin{equation}\tag{A.11}
\widetilde{\mathcal{R}}(U)=(S^{-1}\vec{\Omega})_1\equiv \sum\limits_{k=1}^6 (S^{-1})_{1k}\Omega_k,
\end{equation}
where $S$ is given by (\ref{def-s}) and $\vec{\Omega}$ is a vector of nonlocal variables defined by the relations
\begin{equation}\label{omegaeq}\tag{A.12}
\begin{aligned}
\vec{\Omega}_x&=SM_1 U,\quad &
\vec{\Omega}_t&=SM_2 U.
\end{aligned}
%\end{align*}
\end{equation}
Furthermore, $\widetilde{\mathcal{R}}$ is the inverse of $\mathcal{R}_1$ written in the Guthrie form (\ref{rogut}),
that is, modulo arbitrary integration constants arising from the definition of $W_i$ and $\vec{\Omega}$ we have that
\[
\widetilde{\mathcal{R}}(\mathcal{R}_1(U))= \mathcal{R}_1(\widetilde{\mathcal{R}}(U))=U.
\]
\end{propA}

Upon restating the above result for the recursion operators written in pseudodifferential form we readily arrive at Theorem \ref{t1}.\looseness=-1

\end{document}